\documentclass[11pt]{article}
\input epsf.tex
\newdimen\SaveWidth \SaveWidth=\textwidth
\newdimen\SaveHeight \SaveHeight=\textheight
\advance\SaveWidth by -\textwidth
\advance\SaveHeight by -\textheight
\divide\SaveWidth by 2
\divide\SaveHeight by 2
\advance\hoffset by \SaveWidth
\advance\voffset by \SaveHeight

\def\etal{{\it et al.}}

\def\abs#1{\left| #1\right|}
\def\sgn{\mathop{\rm sgn}}
\def\etmiss{\slashchar{E}_T}
\def\fb{{\rm fb}}

\def\tell{{\tilde\ell}}
\def\ttau{{\tilde\tau}}
\def\fbi{{\rm fb}^{-1}}
\def\Meff{M_{\rm eff}}

\def\lsp{{\tilde\chi_1^0}}

\def\GeV{{\rm GeV}}

\def\tchi{\tilde\chi}
\def\tg{\tilde g}
\def\tq{\tilde q}

\let\badcite=\cite
\def\cite{~\badcite}

\def\slashchar#1{\setbox0=\hbox{$#1$}           
   \dimen0=\wd0                                 
   \setbox1=\hbox{/} \dimen1=\wd1               
   \ifdim\dimen0>\dimen1                        
      \rlap{\hbox to \dimen0{\hfil/\hfil}}      
      #1                                        
   \else                                        
      \rlap{\hbox to \dimen1{\hfil$#1$\hfil}}   
      /                                         
   \fi}                                         %

\catcode`@=11
\newdimen\vbigd@men                             

\def\vbig#1#2{{\vbigd@men=#2\divide\vbigd@men by 2%
   \hbox{$\left#1\vbox to \vbigd@men{}\right.\n@space$}}}
\catcode`@=12

\catcode`@=11
\def\citenum#1{\csname b@#1\endcsname}
\catcode`@=12

\def\dofig#1#2{\centerline{\epsfxsize=#1\epsfbox{#2}}}
\begin{document}
\begin{titlepage}
\rightline{LBNL-43048}
\rightline{BNL-HET-99/10}
\rightline{ATL-COM-PHYS-99-018}

\bigskip\bigskip

\begin{center}{\Large\bf\boldmath
Measurements in SUGRA Models \\
\medskip
with Large $\tan \beta$ at LHC\footnotemark}
\end{center}
\footnotetext{This work was supported in part by the Director, Office of Science,
 Office of Basic Energy Research, Division of High Energy
Physics of the U.S. Department of Energy under Contracts
DE-AC03-76SF00098 and DE-AC02-98CH10886.}
\bigskip
\centerline{\bf I. Hinchliffe$^a$ and F.E. Paige$^b$
}
\centerline{$^a${\it Lawrence Berkeley National Laboratory, Berkeley, CA}}
\centerline{$^b${\it Brookhaven National Laboratory, Upton, NY}}
\bigskip

\begin{abstract}
We present an example  of a scenario of particle
production and decay in supersymmetry models in which the
supersymmetry breaking is transmitted to the observable world via
gravitational interactions. The case is chosen so that there is a
large production of tau leptons in the final state. It is
characteristic of large $\tan\beta$ in that decays into muons and
electrons may be suppressed. It is shown that  hadronic tau decays can
be used to reconstruct final states.

\bigskip

\end{abstract}
\end{titlepage}

\tableofcontents

\newpage
\section{Introduction}
\label{sec:intro}

If supersymmetry (SUSY) exists at the electroweak scale, then
gluinos and squarks will be copiously produced in pairs at the LHC and
will decay via cascades involving other SUSY particles. The
characteristics of these decays and hence of the signals that will
be observed and the measurements that will be made depend upon the
actual SUSY model and in particular on the pattern of supersymmetry
breaking.
Previous, detailed studies of signals for SUSY at the
LHC\cite{hinch,p1,rest,fabiola} have used the SUGRA 
model\cite{SUGRA,SUGRArev}, 
in which the supersymmetry breaking is transmitted to the sector of the 
theory containing the Standard Model particles and their superpartners 
via gravitational interactions. The minimal version of this model has
just four parameters plus a sign.
The lightest supersymmetric particle ($\lsp$) has a mass of order
100~GeV, is stable, is produced in the decay of every other supersymmetric 
particle and  is  neutral and therefore escapes the detector. The
strong production cross sections and the characteristic
signals of events with multiple jets plus missing energy $\etmiss$ or with
like-sign dileptons $\ell^\pm\ell^\pm$ plus
$\etmiss$\cite{BCPT} enable SUSY to be extracted trivially
from Standard Model backgrounds. Characteristic signals were
identified that can be exploited to determine, with great precision,
the fundamental
parameters of these model over the whole of its parameter space.
Variants of this model where 
R-Parity is broken\cite{hall} and the  $\lsp$ is
unstable have also been discussed\cite{rparity}.

These models have characteristic final states depending upon their
parameters. The next to lightest neutral gaugino $\tchi_2^0$ is
produced  in the decays of squarks and gluinos which
themselves may be produced copiously at the LHC. The decay of 
$\tchi_2^0$ then provides a tag from which the detailed analysis of 
supersymmetric events can begin. The dominant decay is usually either
$\tchi_2^0\to h \lsp$ or $\tchi_2^0\to \ell^+\ell^- \lsp$, which can
proceed directly or via the two step decay 
$\tchi_2^0\to \ell^+\tilde{\ell^-}\to \ell^+\ell^- \lsp$. The latter
leads to events with isolated leptons. Both of these characteristic
features
have been explored in some detail in previous
studies\cite{p1,rest,fabiola}.

In the previous cases the smuon, selectron and stau were essentially
degenerate. At larger values of $\tan\beta$, this degeneracy is lifted
and the $\ttau_1$ becomes the lightest slepton. If $m_{1/2}$ is small
enough, then the two-body decays $\tchi_2^0 \to \lsp h$, $\lsp Z$ will
not be allowed, and if $m_0$ is large enough, then $\tchi_2^0 \to
\tell_R\ell$ will also not be allowed. Then for large enough
$\tan\beta$ the only allowed two-body decays are $\tchi_2^0\to
\tau^\pm\tilde\tau^\mp \to \tau^+\tau^- \lsp$. In such cases,
tau decays are dominant, and final states involving tau's must be
used.

The simulation in this paper is based on the implementation of the
minimal SUGRA model in ISAJET\cite{ISAJET}. We use $m_0=m_{1/2}=200$
GeV, $\tan\beta=45$, $A_0=0$ and $\sgn\mu=-1$. The mass spectrum for
this case is shown in Table~\ref{mass-table}. The only allowed
two-body decay of $\tchi_2^0$ is into $\ttau_1\tau$, so it has a
branching ratio of more than 99\%.

\begin{table}[t]
\caption{Masses of the superpartners, in GeV, for the case being studied.
Note that the first
and second generation squarks and sleptons are degenerate and so are 
not listed separately. \label{mass-table}} 
\begin{center}
\begin{tabular}{cccc} \hline \hline 
Sparticle  & mass \qquad &Sparticle & mass\\ \hline
$\widetilde g$  & 540 && \\
 $\widetilde \chi_1^\pm$  & 151 & 
$\widetilde \chi_2^\pm$        &  305 \\
 $\widetilde \chi_1^0$          &  81 & 
$\widetilde \chi_2^0$          &  152   \\
 $\widetilde \chi_3^0$          &  285 &
$\widetilde \chi_4^0$          &  303  \\
$\widetilde u_L$               &  511 & 
$\widetilde u_R$               &  498   \\
$\widetilde d_L$               &  517 & 
$\widetilde d_R$               &  498  \\
$\widetilde t_1$               &  366 & 
$\widetilde t_2$               &  518  \\
$\widetilde b_1$               &   391 &  
$\widetilde b_2$               &  480 \\
$\widetilde e_L$               &  250 & 
$\widetilde e_R$               &  219  \\
$\widetilde \nu_e$             &  237 &
$\widetilde \tau_2$            &  258  \\
$\widetilde \tau_1$            &  132 &  
$\widetilde \nu_\tau$          &    217\\
$h^0$                          &  112 &
$H^0$                          & 157 \\
$A^0$                          &  157
& $H^\pm$                        & 182  \\
\hline \hline
\end{tabular}
\end{center}
\end{table}

The total production cross-section for this model is 99~pb at the LHC. The
rates are dominated by the production of $\tilde{g} \tilde{g}$ and
$\tilde{g}\tilde{q}$ final states. Interesting decays include:

\begin{itemize}
\item $BR(\tchi_2^0\to \tau \tilde{\tau_1})=99.9$\%; $BR(\tchi_1^+\to
  \nu_{\tau} 
\tilde{\tau_1})=99.9$\%
\item  $BR(\tchi_3\to \tchi_2^0 Z)=$13\%;   $BR(\tchi_3\to 
\tau \tilde{\tau_1})=21$\%;
\item $BR(\tilde{g}\to b\tilde{b_1})=55$\%; 
$BR(\tilde{g}\to b\tilde{b_2})=10$\%;      
\item $BR(\tilde{g}\to q_R\tilde{q_l})=3$\%; $BR(\tilde{g}\to
  q_r\tilde{q_r})=5.7$\%;
\item $BR(\tilde{q_L}\to  \tchi_2^0 q)=30$\%;   $BR(\tilde{q_R}\to  \tchi_1^0 q)=97$\%;
\end{itemize}
Here $q$ refers to a light quark.

        All the analyses presented here are based on
ISAJET~7.37\cite{ISAJET} and a simple detector simulation.  600K
 signal events were generated which would correspond to 6 fb$^{-1}$ of 
 integrated luminosity. The Standard Model
background samples contained 250K events for each of $t \bar t$, $WZ$
with $W \to e\nu,\mu\nu,\tau\nu$, and $Zj$ with $Z \to
\nu\bar\nu,\tau\tau$, and 5000K QCD jets (including $g$, $u$, $d$,
$s$, $c$, and $b$) divided among five bins covering $50 < P_T <
2400\,\GeV$. Fluctuations on the histograms reflect the generated
statistics.  On many of the plots that we show, very few Standard
Model background events survive the cuts and the corresponding
fluctuations are large, but in all cases we can be confident that the
signal is much larger than the residual background.  The cuts that we
chosen have not been optimized, but rather have been chosen to get
background free samples.

        The detector response is parameterized by Gaussian resolutions
characteristic of the ATLAS\cite{ATLAS} detector without any tails.
All energy and momenta are measured in GeV.  In the central region of
rapidity we take separate resolutions for the electromagnetic (EMCAL)
and hadronic (HCAL) calorimeters, while the forward region uses a
common resolution:
\begin{eqnarray*}
{\rm EMCAL} &\quad& \phantom{0}10\%/\sqrt{E} \oplus 1\%, 
\quad |\eta|\, <3 \,\nonumber\\
{\rm HCAL}  &\quad& \phantom{0}50\%/\sqrt{E} \oplus 3\%, 
\quad |\eta|\, < 3\,\nonumber\\
{\rm FCAL}  &\quad& 100\%/\sqrt{E} \oplus 7\%, 
\quad |\eta|\, > 3\,\nonumber
\end{eqnarray*} 
A uniform segmentation $\Delta\eta = \Delta\phi = 0.1$ is used with no
transverse shower spreading. Both ATLAS\cite{ATLAS} and CMS\cite{CMS}
have finer segmentation over most of the rapidity range, but the
neglect of shower spreading is unrealistic, especially for the forward
calorimeter. Missing transverse energy is calculated by taking the
magnitude of the vector sum of the transverse energy deposited in in
the calorimeter cells. An oversimplified parameterization of the muon
momentum resolution of the ATLAS detector --- including a both the
inner tracker and the muon system measurements --- is used, {\it viz}
$$
\delta P_T/P_T = \sqrt{0.016^2+(0.0011P_T)^2}
$$
For electrons we use a momentum resolution obtained by combining the
electromagnetic calorimeter resolution above with a tracking
resolution of the form
$$
\delta P_T/P_T= \left(1+{0.4\over(3-\abs{\eta})^3}\right)
\sqrt{(0.0004P_T)^2+0.0001}
$$
This provides a slight improvement over the calorimeter alone.

        Jets are found using GETJET\cite{ISAJET} with a simple
fixed-cone algorithm. The jet multiplicity in SUSY events is rather
large, so we will use a cone size of 
$$
R = \sqrt{(\Delta\eta)^2+(\Delta\phi)^2} = 0.4
$$ 
unless otherwise stated. Jets are required to have at least $P_T >
20\,\GeV$; more stringent cuts are often used.  All leptons are
required to be isolated and have some minimum $P_T$ and $\abs{\eta}<
2.5$, consistent with the coverage of the central tracker and muon
system. An isolation requirement that no more than 10~GeV of
additional $E_T$ be present in a cone of radius $R = 0.2$ around the
lepton is used to reject leptons from $b$-jets and $c$-jets. In
addition to these kinematic cuts a lepton ($e$ or $\mu$) efficiency of
90\% and a $b$-tagging efficiency of 60\% is assumed\cite{ATLAS}. 

As $\tau$'s are a crucial part of this analysis, they require special
treatment. We concentrate on hadronic tau decays, since for leptonic
decays the origin of the lepton is not clear and the visible lepton in
general carries only a small fraction of the true tau momentum.
Using the fast simulation, we first identify the hadronic taus by
searching the reconstructed jet list for jets with $P_T>20$ GeV and
$\abs{\eta}<2.5$. We then compare these jets with the generated tau
momenta and assign them to a reconstructed tau if $E_{\tau}> 0.8
E_{jet}$ and the center of the jet and the tau are separated by
$\Delta R < 0.4$. 

We then rely on a full simulation\cite{ianhtau} of $Z+{\rm jet}$
events with $Z \to \tau\tau$. Events were generated with PYTHIA \cite{pythia} and
passed through the ATLAS GEANT simulation (DICE) and reconstruction
(ATRECON) programs \cite{dice}. The charged particles were reconstructed with the
tracking and the photons with the calorimeter. Cuts were then applied
to the invariant mass and isolation of the reconstructed taus.
These cuts produce a rejection factor against QCD jets of a factor of
15 and accept 41\% of the hadronic tau decays. We apply these results
to the hadronic tau's identified in our fast simulation on a
probabilistic basis. The  accepted hadronic decays are
assumed to be measured using the resolution from the full simulation,
while the ones not accepted are put back into the jet list. Fake
$\tau$'s are made by reassigning jets with the appropriate
probability. The full simulation also indicates that the tau charge
is correctly identified 92\% of the time. We include this factor in
our fast $\tau$ reconstruction and assign the fake tau's to either
sign with equal probability.  For cases where the $\tau\tau$ invariant
mass is to be measured, the generated $\tau\tau$ invariant mass is
smeared with a resolution derived from the full simulation, i.e., a
Gaussian with a peak at $M=0.66M_{\tau\tau}$ and $\sigma /M=0.12$. In
cases where the measured  momentum of the  $\tau$ decay products 
is needed, the measured jet energy is used. 

        Results are presented for an integrated luminosity of
$10\,\fb^{-1}$, corresponding to one year of running at $10^{33}\,{\rm
cm^{-2}s^{-1}}$; pile up has not been included. We will occasionally
comment on the cases where the full design luminosity of the LHC, {\it
i.e.}\ $10^{34}\,{\rm cm^{-2}s^{-1}}$, will be needed to complete the
studies. For many of the histograms shown, a single event can give
rise to more than one entry due to different possible combinations.
When this occurs, all combinations are included.

The rest of this paper is organized as follows. We first illustrate
how measurement of the $\tau\tau$ final state can be used to infer
information on the masses of the staus. We then use this final state
in conjunction with $b$-jets to reconstruct gluinos and bottom
squarks. Methods for extracting information on light squarks are then
shown and the dilepton mass distribution is used to give information
on the masses $\tchi_4^0$. Finally we show how this information can be 
combined to constrain the underlying model parameters.

\section{Effective mass distribution}
\label{sec:meff}

\begin{figure}[t]
\dofig{3.20in}{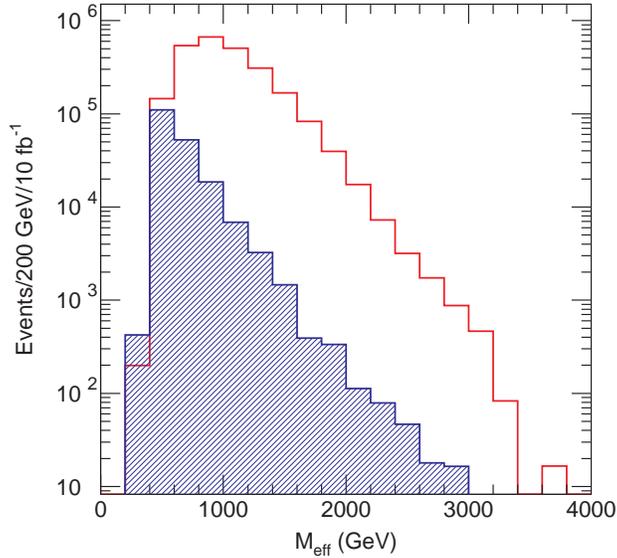}
\caption{SUSY signal (open histogram) and Standard Model backgrounds
  (filled histogram).\label{point6meff}} 
\end{figure}

        The first step in the search for new physics is to discover a
deviation from the Standard 
Model and to estimate the mass scale associated with it. SUSY
production at the LHC is dominated by gluinos and squarks, which decay
into multiple jets plus missing energy. A variable which is sensitive
to inclusive gluino and squark decays is the effective mass $\Meff$,
defined as the scalar sum of the $P_T$'s of the four hardest jets and
the missing transverse energy $\etmiss$,
$$
\Meff = p_{T,1} + p_{T,2} + p_{T,3} + p_{T,4} + \etmiss\,.
$$
Here the jet $P_T$'s have been ordered such that $p_{T,1}$ is the
transverse momentum of the leading jet.
The Standard Model backgrounds tend to have smaller $\etmiss$, fewer
jets and a lower jet multiplicity.  In addition,
since a major source of $\etmiss$ is weak decays, large $\etmiss$
events in the Standard Model
tend to have the missing energy associated with leptons.  To
suppress these backgrounds, the following cuts were made:  
\begin{itemize}
\item   $\etmiss > 100\,\GeV$;
\item   $\ge4$ jets with $P_T > 50\,\GeV$ and  $p_{T,1} > 100\,\GeV$;
\item   Transverse sphericity $S_T > 0.2$;
\item   No $\mu$ or isolated $e$ with $P_T > 20\,\GeV$ and $|\eta|<2.5$; 
\item   $\etmiss > 0.2 \Meff$.
\end{itemize}
Note that some of these jets could result from  hadronic tau decays.
With these cuts and the idealized detector assumed here, the signal is
much larger than the Standard Model backgrounds for large $\Meff$, as
is illustrated in Figure~\ref{point6meff}. Thus, the discovery
strategy developed for low $\tan\beta$\cite{hinch} also works for
this case.
As demonstrated in more detail elsewhere\cite{hinch} the shape of
this effective mass distribution can be used to estimate the masses of 
the SUSY particles that are most copiously produced; here quarks and gluinos.

\section{Tau-Tau invariant mass}
\label{sec:tau}

As can be seen from the decays listed above we expect significant
prodution of $\tchi_2^0$ and hence of tau pairs from the decay of
 $\tilde{q_L}$ We require that the events contain at least two jets that are
indentified as hadronic tau decays using the above algorithm.  In
addition, the following cuts are applied:
\begin{itemize}
\item   $\etmiss > 100\,\GeV$;
\item   at least four jets with $P_T > 50\,\GeV$ and  at least one jet 
$p_{T,1} > 100\,\GeV$;
\item   $\Meff > 500\,\GeV$;
\item   $\etmiss > 0.2 \Meff$.
\end{itemize}
Again, some of these jets could result from  hadronic tau decays.

\begin{figure}[t]
\dofig{3.20in}{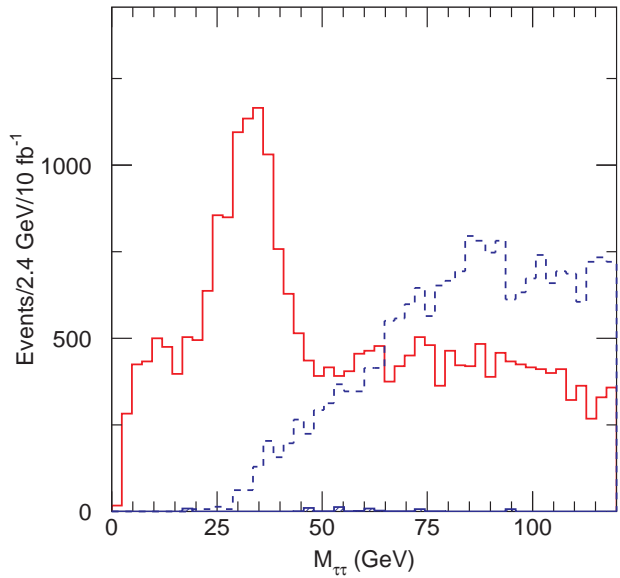}
\caption{Reconstructed $\tau\tau$ mass distribution. All
  combinations of tau pairs are shown irrespective of the charge. The
  dashed histogram shows the comibination of one real tau and one fake 
  tau. The actual experiement would observe the sum of the two
  histograms.
The background from Standard Model processes is very small and is not shown.
\label{mtautau}} 
\end{figure}

\begin{figure}[t]
\dofig{3.20in}{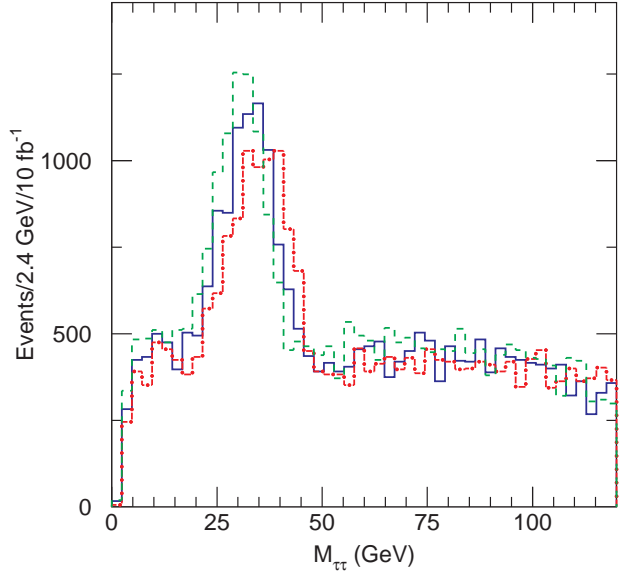}
\caption{Reconstructed $\tau\tau$ mass distribution  showing the 
  effect of rescaling the generated tau-tau invarient mass
  distribution by  $\pm 7.5$\%. \label{mtautau1}}
\end{figure}

\begin{figure}[t]
\dofig{3.20in}{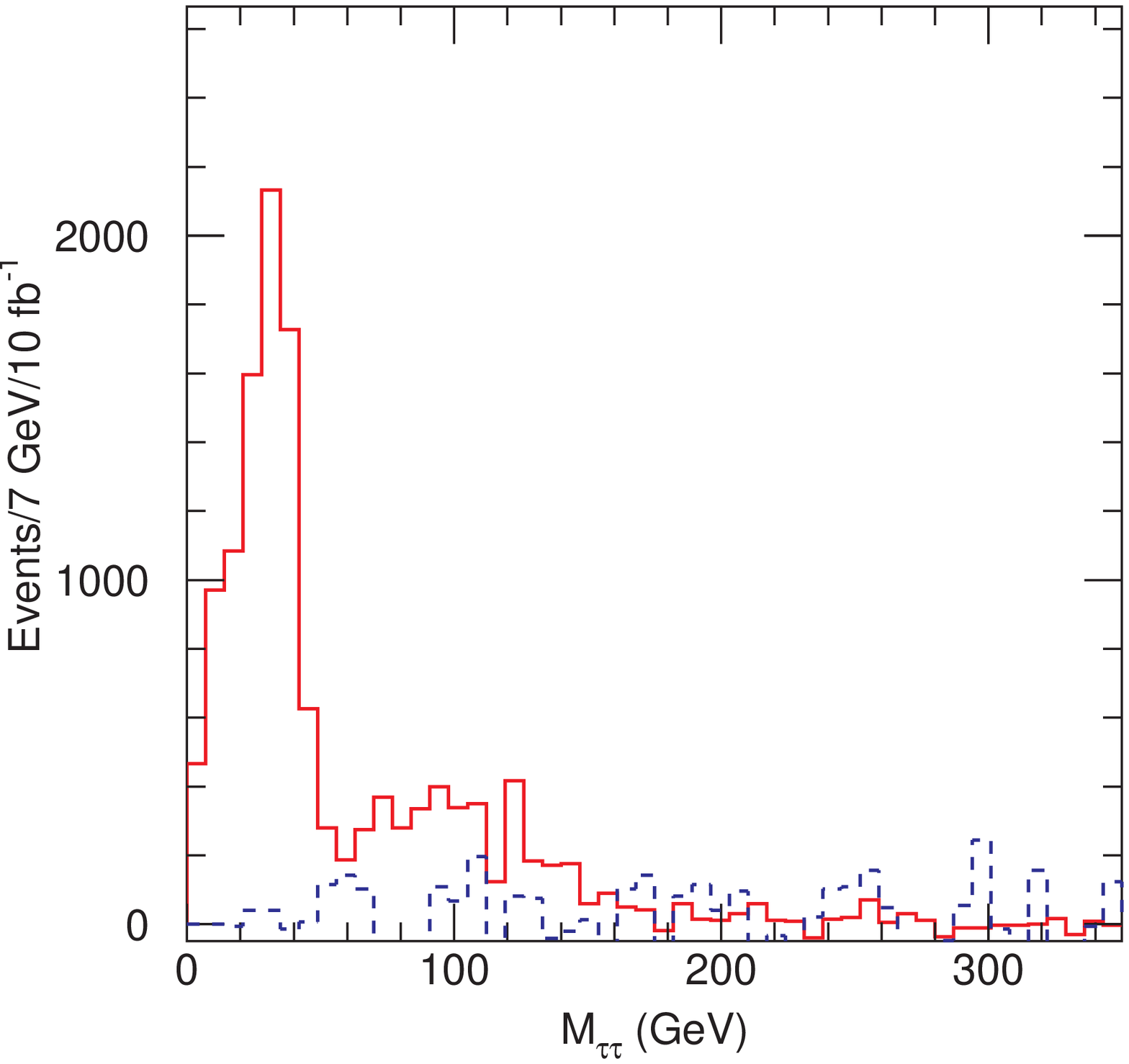}
\caption{Reconstructed $\tau^+\tau^- - \tau^\pm\tau^\pm$ mass
distribution. The dashed line shows the fake-real
background. The flucuations are slightly larger than the true
statistics. \label{subtracted}}
\end{figure}

We then search for taus that decay hadronically using the algrotihm
discussed above. The
reconstructed $\tau\tau$ invarient mass distribution is shown in
 Figure~\ref{mtautau}; all combinations of tau charges are shown in
 this Figure. It can be seen from this distribution that
 there is a clear structure. There is considerable background from
 combinations where one of the identified tau jets is from a tau and the
 other is from a misidentified jet.
 The invarient mass distribution of
 these pairs is also shown in Figure~\ref{mtautau}; it is rather
 featureless. The tau algorthm has not been optimized so this
 background could well have been overestimated.
 The background from events where both taus are
 misidentified jets and the Standard Model background are both
 negligible. The position of the peak in this
 mass distribution enables one to infer the position of the end
 point arising from the decay chain $\tchi_2^0\to \tau \tilde\tau_1 \to 
 \tchi_1^0 \tau \tau$:
$$
M_{\tau\tau}^{\rm max} = M_{\tilde\chi_2^0}
\sqrt{1-{M_{\tilde\ell}^2 \over M_{\tilde\chi_2^0}^2}}
\sqrt{1-{M_{\lsp}^2 \over M_{\tilde\ell}^2}} = 59.6\,\GeV.
$$ 

In order to estimate the precision with which this endpoint can be
determined, the generated tau-tau invarient mass distribution was
shifted by  $\pm 7.5$\% from its nominal value. The effect on the
reconstructed $\tau\tau$ mass distribution is shown in
Figure~\ref{mtautau1}. These cases can clearly be distinguished. The
actual precision that can be obtained on the position of this end
point requires a more detailed study. Tau decays are well understood;
the problem is to determine the effects of the detector resolutions
and the cuts. For the purposes of extracting
parameters below, we will assume an uncertainty of 5\% can be achieved.

There are some events beyond this edge as can be seen by looking at
the subtracted distribution $\tau^+\tau^- - \tau^-\tau^- -
\tau^+\tau^+$ shown in Figure~\ref{subtracted}.  This subtraction also
elliminates the background from fake taus because their charges are
not correlated. Here the excess extends to $\sim 150$ GeV and is due to
$\chi_3$ and $\chi_4$ decays. This can be confirmed by the large $Z$
signal (see below). The fluctuations in this histogram reflect the
generated statistics, which correspond to about $6\,\fbi$; three years
at low luminosity would make this high-mass signal much clearer.

\section{Reconstruction of $\tilde{g}\to b \tilde{b} \to b \tchi_2^0
b\to bb \tau^+\tau^-\lsp$} \label{sec:gluino}

\begin{figure}[t]
\dofig{3.20in}{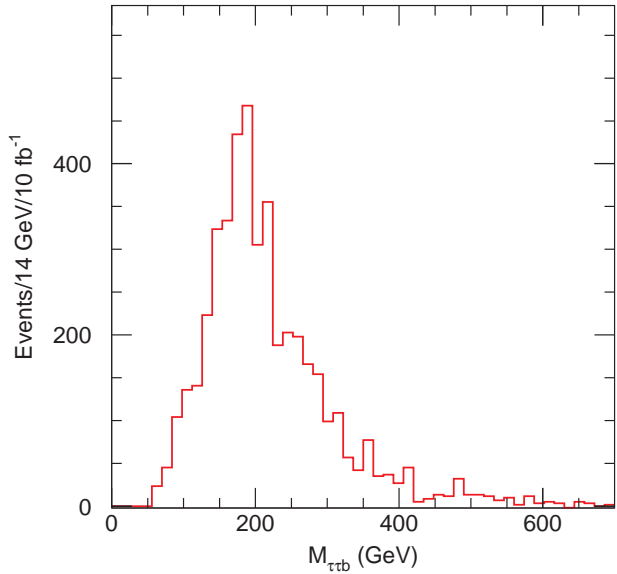}
\caption{Reconstructed $\tau\tau+{\rm jet}$ mass
distribution  where the jet is tagged as a $b$-jet. The background
from Standard Model processes is negligible.
\label{sbot}} 
\end{figure}

The event sample of the previous section is used in an attempt to
reconstruct squarks and gluinos. We concentrate here on final states with
$b$ quarks as these have the larger branching ratios and less
combinatorial background.
In addition to the previous cuts, we
require a tagged $b$-jet with $P_T> 25 $ GeV; this jet
could be one of the ones in the previous selection. Events are
selected that have reconstructed tau pairs with invariant mass within
$\pm 10$ GeV of peak in Figure~\ref{mtautau}, and the invariant mass of
the tau pair and the $b$-jet is formed. This mass distribution is shown
in Figure~\ref{sbot}.  The sign subtracted distribution corresponding
to $\tau^+\tau^- - \tau^-\tau^- - \tau^+\tau^+$ is used to
reduce combinatorial background.  There should should be an edge at
$\sim m_{\tilde{b}_1}-m_{\tchi_1}=310$~GeV. The edge is not sharp --- 3
particles are lost, two $\nu_{\tau}$'s and the $\lsp$. In
addition the distribution is contaminated by decays from $\tchi_3^0$ and
$\tchi_4^0$. The structure is not clear, but is well distinguished from
that resulting from the case where the $b$-jet is replaced by a light
quark jet, shown in Figure~\ref{p6left}.

\begin{figure}[t]
\dofig{3.20in}{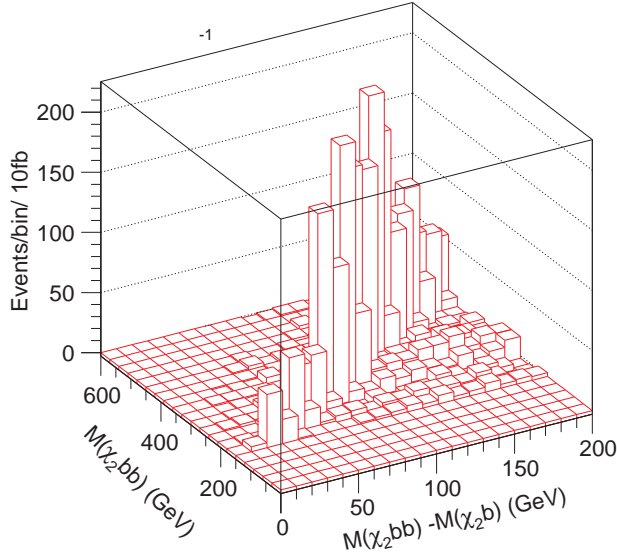}
\caption{Lego plot showing the reconstructed masses  $m(\tchi_2^0
b)$ and  $m(\tchi_2^0 bb) - m(\tchi_2^0 b)$. \label{p6lego}}
\end{figure}

Further information can be obtained by applying a partial
reconstruction technique. This was developed in Ref.~\citenum{hinch} 
(so called ``Point~3'') where the
decay chain  $\tilde{g}\to b \tilde{b} \to  b b \tchi_2^0  b \to  b b 
\ell^+\ell^- \lsp$ was fully reconstructed as follows. 
If the mass of the lepton
pair is near its maximum value, then in the rest frame of
$\tchi_2^0$ both $\tchi_1$ and the $\ell^+\ell^-$ pair are forced to be at
rest. The momentum of $\tchi_2^0$ in the laboratory frame is then
determined  
$$
\vec P_{\tchi_2^0}=\left(1+M_{\lsp}/M_{\ell^+\ell^-}\right) 
\vec P_{\ell^+\ell^-}\,.
$$ 
where  $P_{\ell^+\ell^-}$ is the momentum of the dilepton system. 
The $\tchi_2^0$ can then be combined with b jets to reconstruct the decay
 chain. A clear correlation between  the masses of the $b\tchi_2^0$ and 
 $bb \tchi_2^0$ systems was observed allowing both the gluino and
 sbottom masses to be determined if the mass of $\lsp$ was assumed.
The inferred mass difference $m_{\tilde{g}}-m_{\tilde{b}}$ was found
to be insensitive to assumed $\lsp$ mass.

\begin{figure}[t]
\dofig{3.20in}{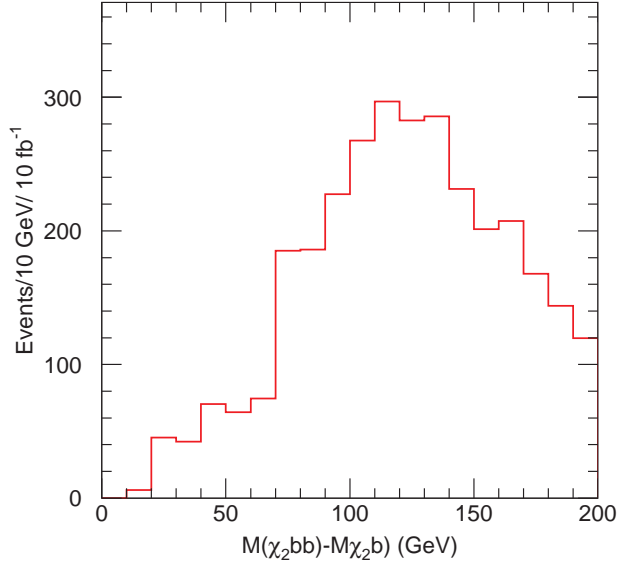}
\caption{Projection of Figure~\protect\ref{p6lego} onto the
$m(\tchi_2^0 bb)- m(\tchi_2^0 b)$ axis. \label{p6diff}}
\end{figure}

\begin{figure}[t]
\dofig{3.20in}{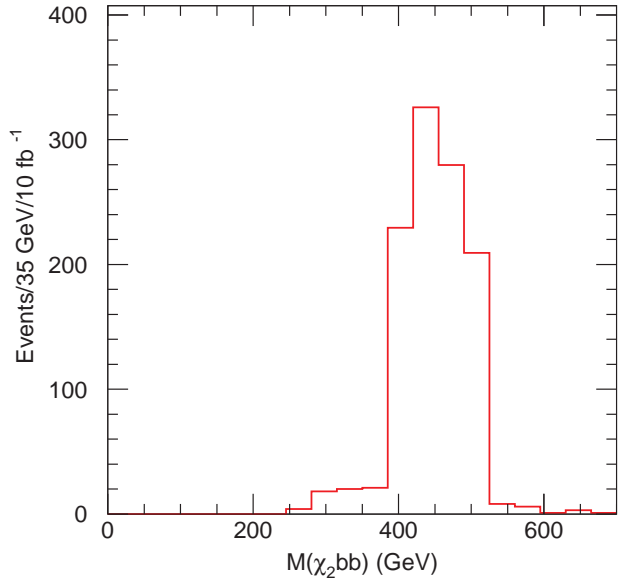}
\caption{Projection of Figure~\protect\ref{p6lego} onto the $m(\tchi_2^0 bb)$
  axis with the requirement that  $100\,\GeV < m(\tchi_2^0 bb) - m(\tchi_2^0
  b)< 140\,\GeV$.  \label{p6proj}}
\end{figure}

In the case of interest here the situation is more complicated. First, there
is an extra step in the decay chain {\it i.e.}\  $\tilde{g}\to b
\tilde{b} \to  b b \tchi_2^0  b \to b b \tau \ttau \to  b b 
\tau^+\tau^- \lsp$. So that even if the events could be selected 
such  that the
$\tau\tau$ invariant mass was at the kinematic limit, $\lsp$ would not 
 be at rest in the $\tchi_2^0$ rest frame, and  the inferred
 $\tchi_2^0$ momenta would not be correct.  This was the case at
 ``Point~5'' \cite{hinch} 
 where the method was applied to the decay chain $\tilde{q} \to q 
\tchi_2^0 \to q \mu \tilde{\mu} \to q \mu^+\mu^- \lsp$ and nevertheless, a
mass peak was reconstructed in that case.
Second, the momentum of the $\tau\tau$
system cannot be measured owing to the lost energy from neutrinos.
Despite these problems the method is still effective as is now
demonstrated. We select events with reconstructed $\tau\tau$ mass in
the range
$$
40 \, \GeV \, < m_{\tau\tau} <  60\,\GeV
$$ 
and infer the momentum of $\tchi_2^0$ from
the measured momentum $P_{\tau^+\tau^-}$ of the $\tau\tau$ system
assuming the nominal value of $M_{\lsp}$ .
$$
\vec P_{\tchi_2^0}=\left(1+M_{\lsp}/M_{\tau^+\tau^-}\right) 
\vec P_{\tau^+\tau^-}\,.
$$ 
This momentum is then combined with that of two measured b-jets each
required to have $P_T> 25$~GeV and the mass of the $\tchi_2^0 b$ and
$\tchi_2^0 bb$ systems computed. 
Figure~\ref{p6lego} shows the correlation $m(\tchi_2^0
b)$ vs.\ $(m(\tchi_2^0 bb) - m(\tchi_2^0 b))$ in  a lego plot. The subtracted
distribution corresponding to  $\tau^+\tau^- - \tau^-\tau^- -
\tau^+\tau^+$ is used to reduce the background.
There is a clear peak in this plot.
The projection of this plot onto the $m(\tchi_2^0 bb) - m(\tchi_2^0 b)$ axis 
is shown in Figure~\ref{p6diff} which shows a peak at 120 GeV,
somewhat below the true mass difference of 150~GeV. If a selection of
events with  120 GeV $< m(\tchi_2^0 bb) - m(\tchi_2^0 b)< 140$ GeV is made
and Figure~\ref{p6lego} projected onto the  $m(\tchi_2^0 bb)$ axis the
result is shown in Figure~\ref{p6proj}. A fairly sharp peak results at a
value somewhat below the gluino mass of 540~GeV. This displacement to
lower values is due to two effects; jet energy is lost out of the
clustering cone and carried off by neutrinos is semileptonic bottom and 
charm decays. We have not recalibrated the $b-$jet energy scale to take 
account of these effects.

\section{Light Squarks} 

\begin{figure}[t]
\dofig{3.20in}{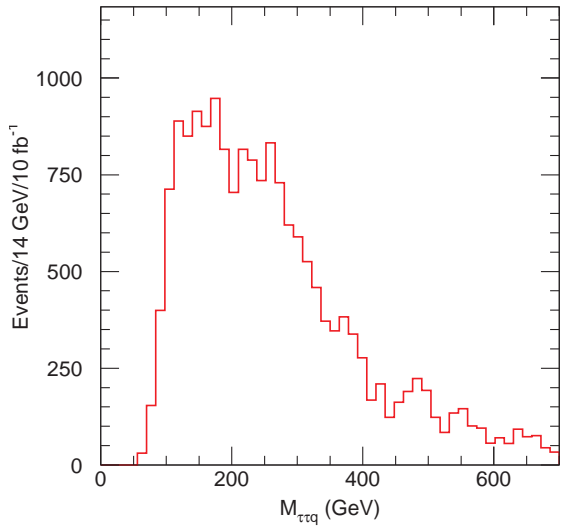}
\caption{Reconstructed $\tau\tau + {\rm jet}$ mass
distribution at for light quark jets. \label{p6left}}
\end{figure}

We now attempt to find evidence for the decay chain $\tq_L \to q
\tchi_2^0 \to q \tilde{\tau} \tau \to q \tau\tau \lsp$. The rates are
not large due to the small branching ratio for the first step, and we can 
expect considerable combinatorial background from QCD radiation of
light quark and gluon jets.
The event sample of section~\ref{sec:tau} is used. In addition we
require the presence of a non $b$-jet with $P_T> 25 $ GeV.  Events are
selected that have reconstructed tau pairs with invarient mass within
$\pm 10\,\GeV$ of peak in Figure~\ref{mtautau}, and the invariant mass of
the tau pair and the jet is formed. This mass distribution is shown
in Figure~\ref{p6left}.  The sign subtracted distribution
corresponding to $\tau^+\tau^- - \tau^-\tau^- - \tau^+\tau^+$ is
used as it reduces combinatorical background.  There should should be
an edge at $\sim m_{\tilde{q_l}}-m_{\tchi_1}\sim 400\,\GeV$. The edge is not
sharp --- two $\nu_{\tau}$'s and the $\lsp$ are all lost.  In addtion
the distribtution is contaminated by decays from $\tchi_3^0$ and
$\tchi_4^0$. While this distribution is clealy distinct from that
shown above where b-jets were used, more work is needed to establish
that this could  be used to infer information on the light squark mass.

\section{Extraction of $\tq_R$}
\label{sec:qright}

This analysis is based on the fact that $\tq_R \to q \lsp$ is
dominant, so $\tq_R \tq_R$ pair production gives a pair of hard jets
and large missing energy. There is no kinematic endpoint, but the
$P_T$ of the jets provides a measure of the squark mass \cite{p1}. 
The following
cuts were made:
\begin{itemize}
\item   $\etmiss > 200\,\GeV$
\item   2  jets with $P_T > 150\,\GeV$ 
\item   No other jet with   $p_{T} > 25\,\GeV$
\item   Transverse sphericity $S_T > 0.2$
\item   $\etmiss > 0.2 \Meff$
\item   No leptons, No b-jets, No tau jets
\end{itemize}

\begin{figure}[t]
\dofig{3.20in}{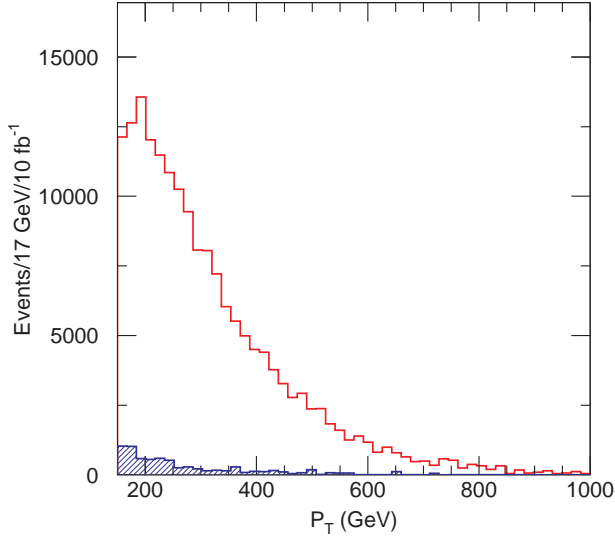}
\caption{Transverse momentum distribution for jets passing the
  selection described in section~\ref{sec:qright}. The Standard Model
  background is shown as the filled histogram.  \label{qright}}
\end{figure}

The transverse momentum distribution of the leading jets is shown in
Figure~\ref{qright}. The error on the mass is limited by the
systematics of understanding the production dynamics and the SUSY
backgrounds. Studies of other cases \cite{p1} have shown that this
distribution should enable a precision of  $\pm 50\,\GeV$ to be reached; it
might be possible to achieve $\pm 25\,\GeV$ in a high statistics study.

\section{Dilepton Final states}
\label{sec:dilep}

While the light gauginos decay almost entirely into $\tau$'s, the
heavy ones can decay via $\tchi_{3,4}^0 \to \tell_{l,r}^\pm \ell^\mp
\to \tchi_{1,2}^0 \ell^+\ell^-$, giving opposite-sign, same-flavor
leptons. The largest combined branching is for $\tchi_4^0 \to
\tell_l^\pm\ell^\mp \to \lsp\ell^+\ell^-$, which gives a dilepton
endpoint at 
$$
M_{\ell\ell}^{\rm max} = \sqrt{(M_{\tchi_4^0}^2 - M_{\tell_L}^2)
(M_{\tell_L}^2 - M_{\lsp}^2)\over M_{\tell_L}^2} = 163.2\,\GeV
$$
There is of course a large background from leptonic $\tau$ decays, but
this can be cancelled statistically by measuring the flavor-subtracted
combination $e^+e^- + \mu^+\mu^- - e^\pm\mu^\mp$ as we now demonstrate. 

\begin{figure}[t]
\dofig{3.20in}{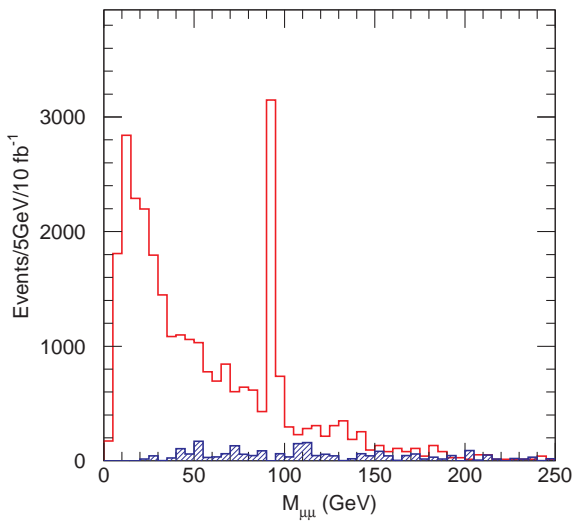}
\caption{Reconstructed $\mu^+\mu^-$ mass
  distribution.\label{mmumu}. The filled histogram shows the Standard 
  Model background.}
\end{figure}

\begin{figure}[t]
\dofig{3.20in}{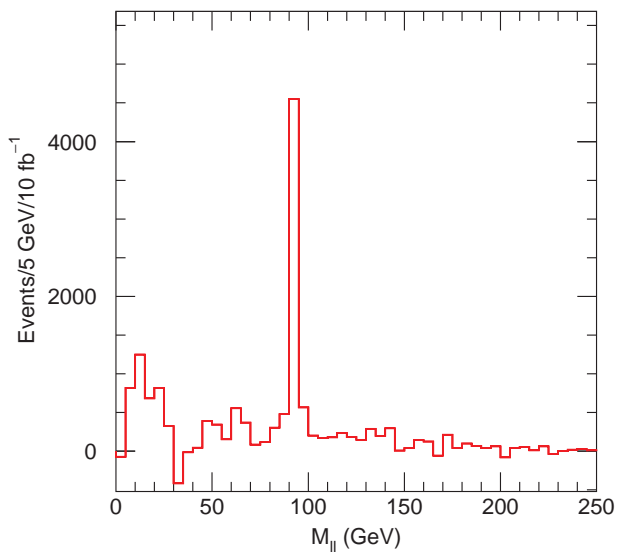}
\caption{Reconstructed $\mu^+\mu^- +e^+e^- - \mu^+e^- - \mu^-e^+$  
mass distribution.\label{mmumusub}}
\end{figure}

Events were selected to have two leptons with $P_T>10\,\GeV$ and
$|\eta|<2.5$ in addition to the jet and $\etmiss$ cuts described
earlier (see Section~\ref{sec:tau}: no tau requirement is applied here). Figure~\ref{mmumu} shows the distribution in the $\mu^+\mu^-$ 
final state. A clear peak from $Z$ decay is visible that results from 
 $\chi_3$ and $\chi_4$ decays. 
 The flavor-subtracted
combination $e^+e^- + \mu^+\mu^- - e^\pm\mu^\mp$ is shown in
Figure~\ref{mmumusub} and shows an excess extending to $\sim 160$
GeV. Unlike the distributions involvin tau final states, this one can
be extrapolated to high luminosity operation which will surely be needed 
to extract a quantitative result from it.

\section{Determining SUSY parameters and Conclusion}
\label{sec:scan}

The presence of the dijet signal of Section~\ref{sec:qright} implies
that that $m_{\tg} > m_{\tq_R}$. Likewise the failure to
observe a dilepton peak implies that $m_{\tilde{e}_R}>m_{\tchi_2^0}$.
These results are used together with the assumed errors on the measured
quantities to fit the model parameters.  
The values of the errors on
$m_{\tilde{g}}-m_{\tilde{b}}$, $m_{\tilde{g}}$, $m_{\tilde{q_r}}$ and
the $\tau\tau$ edge are shown in Table~\ref{tab-fit}.  We do not use
the information from Figures \ref{p6left} and \ref{mmumusub} as we have not
estimated the quantitative information that they could give. Two fits are
shown since the sign of $\mu$ cannot be determined. This is
expected: a change of conventions can replace $\sgn\mu$ with
$\sgn(\tan\beta)$, and $\tan\beta=\pm\infty$ are equivalent.

We assume that the Higgs mass is measured via its decay to two
photons. The error on the Higgs mass is likely to be dominated by the
theoretical 
uncertainty on the higher order corrections; both the one-loop and the
dominant two-loop effects have been calculated and are large. The
present error is probably about $\pm 3\,\GeV$; this might be reduced to
$\pm1\,\GeV$ with much more work. 
 The ultimate limit comes from the
experimental error, about $\pm 0.2\,\GeV$. The effect of reducing this error is only apparent in the error of the 
fitted value of $\tan\beta$ whose error is reduced by approximately a
factor of two if $\pm1\,\GeV$ error on the Higgs mass is used.
 The table shows various assumptions for the errors that might
be achieved. The numbers in the first column are conservative and
will be achieved with the 10 fb$^{-1}$ of integrated luminosity shown
on the figures. The rightmost column is an estimate of what might
ultimately be achievable. We caution the reader that the measurements
involving tau's may not be possible at a luminosity of $10^{34}$
cm$^{-2}$
 sec$^{-1}$ due to pileup effects.

\begin{table}
\caption{Results of the fit for the model parameters. The assumed
errors in GeV on the measured quantities are shown for low and high
luminosity with two different assumptions about how well the $\tq_R$
mass can be extracted from Figure~\ref{qright}. The fitted values of
$m_0$ $m_{1/2}$, $\tan\beta$ and $A_0$ are given for each case for
both signs of $\mu$.  The theoretical plus experimental error on the
light Higgs mass is assumed to be 3 GeV.\label{tab-fit}
}
\begin{center}
\begin{tabular}{|c|c|c|c|c|}
\hline
$\cal L$ & \multicolumn{2}{|c|}{$10\,\fbi$} &
\multicolumn{2}{|c|}{$100\,\fbi$} \\
\hline
$\tau\tau$ edge& 3.0 & 3.0 & 1.2 & 1.2 \\
\hline
$m_{\tilde{g}}-m_{\tilde{b}}$& 20.& 20.& 10.& 10. \\
\hline
$m_{\tilde{g}}$& 60.& 60. &30.& 30.\\
\hline
$m_{\tilde{q_r}}$& 50.& 25.& 25.& 12.\\
\hline
\multicolumn{5}{|c|}{$\mu>0$}\\
\hline
$m_0$ &$232\pm 39$&  $ 228\pm 27$&$   230\pm 30 $&$   227 \pm 29$\\
\hline
$m_{1/2}$&$ 198 \pm 14 $&$ 195 \pm 11 $&$  196 \pm 10 $&$ 195 \pm 9$\\
\hline
$\tan\beta$&$ 42 \pm 7  $&$  43 \pm 6 $&$ 44 \pm 5.5$&$ 44 \pm 5$\\
\hline
$A_0$&$   0 \pm 200  $&$   0 \pm 180 $&$  161 \pm 150$&$ -60 \pm 140$\\
\hline
\multicolumn{5}{|c|}{$\mu<0$}\\
\hline
$m_0$&$ 230 \pm 37 $&$  232 \pm 26 $&$ 230 \pm 26  $&$ 233 \pm 26$\\
\hline
$m_{1/2}$&$  200 \pm 14  $&$ 196 \pm 11   $&$ 198 \pm 7 $&$ 201 \pm 6$\\
\hline
$\tan\beta$&$ 42 \pm 7.3  $&$  42 \pm 7.1  $&$  45 \pm 6.2 $&$ 45 \pm 6.1$\\
\hline
$A_0$& $0 \pm 270  $&$     0 \pm 270 $&$ -100 \pm 210$&$ -150 \pm 200$\\
\hline
\end{tabular}
\end{center}

\end{table}

We can see from the table that, despite the fact that the tau
momenta cannot be measured directly due to the presence of neutrinos in their
decays, we can still expect to infer values of the underlying
parameters with errors of better than 10\%.  Of course these errors
are considerably poorer than those that we expect in cases where taus
do not have to be used\cite{hinch}. Our encouraging result arises
mainly from the very large statistical sample that LHC can produce for 
the case considered.

\section*{Acknowledgements}

This work was supported in part by the Director, Office of Science,
 Office of Basic Energy Research, Division of High Energy
Physics of the U.S. Department of Energy under Contracts
DE-AC03-76SF00098 and DE-AC02-98CH10886.  Accordingly, the U.S.
Government retains a nonexclusive, royalty-free license to publish or
reproduce the published form of this contribution, or allow others to
do so, for U.S. Government purposes.

\end{document}